\RequirePackage{lineno}
\documentclass[aps,prd,twocolumn,superscriptaddress,showpacs,amsmath,amssymb]{revtex4}

\usepackage{comment}
\usepackage{multirow}
\usepackage{dcolumn}
\usepackage{bm}
\usepackage{lineno}
\usepackage{setspace}
\usepackage{graphicx}
\includecomment{printrobustnotes}
\includecomment{printallnotes}
\usepackage{xspace}

\usepackage{dcolumn}
\usepackage{bm}

\newenvironment{cfigure1c}[1][tbp]{\begin{figure*}[#1]\centering}{\end{figure*}}

\newcommand{\note}[1]{\xspace}
\newcommand{\robustnote}[1]{\xspace}
\begin{printrobustnotes}
\renewcommand{\robustnote}[1]{\textbf{\textit{#1}}\xspace}
\end{printrobustnotes}
\begin{printallnotes}
\renewcommand{\note}[1]{\textbf{\textit{#1}}\xspace}
\renewcommand{\robustnote}[1]{\textbf{\textit{#1}}\xspace}
\end{printallnotes}
\newcommand{\ppbar}  {\ensuremath{p\bar{p}}\xspace}
\newcommand{\ttbar}  {\ensuremath{t\bar{t}}\xspace}

\newcommand{\Ht}     {\ensuremath{H_{T}}\xspace}
\newcommand{\et}     {\ensuremath{E_{T}}\xspace}
\newcommand{\etl}     {\ensuremath{E_{T}^{\mathrm{lepton}}}\xspace}
\newcommand{\etm}     {\ensuremath{E_{T}^{\mathrm{muon}}}\xspace}
\newcommand{\etj}     {\ensuremath{E_{T}^{\mathrm{jet}}}\xspace}

\newcommand{\pt}     {\ensuremath{p_{T}}\xspace}
\newcommand{\ptt}     {\ensuremath{p_{T}^{\mathrm{total}}}\xspace}
\newcommand{\ptl}     {\ensuremath{p_{T}^{\mathrm{lepton}}}\xspace}
\newcommand{\ptm}     {\ensuremath{p_{T}^{\mathrm{muon}}}\xspace}

\newcommand{\met}    {\ensuremath{\text{\raisebox{.3ex}{$\not$}}\et}\xspace}

\newcommand{\mtop}    {\ensuremath{\mathrm{M}_{\text{top}}}\xspace}

\newcommand{\mtr}     {\ensuremath{m_{t}^{\text{reco}}}\xspace}
\newcommand{\mttr}     {\ensuremath{m_{t}^{\text{reco(2)}}}\xspace}

\newcommand{\mtnwa}   {\ensuremath{m_{t}^{\text{NWA}}}\xspace}

\newcommand{\mjj}    {\ensuremath{m_{jj}}\xspace}

\newcommand{\mTT}    {\ensuremath{m_{\text{T2}}}\xspace}

\newcommand{\djes}   {\ensuremath{\Delta_{\mathrm{JES}}}\xspace}

\newcommand{\genunit}[2]{\ensuremath{#1~\mathrm{#2}}\xspace}

\newcommand{\gev}[1]    {\genunit{#1}{GeV}}
\newcommand{\pb}[1]     {\genunit{#1}{pb}}

\newcommand{\gevc}[1]   {\ensuremath{#1~\mathrm{GeV}/c}}
\newcommand{\gevcc}[1]  {\ensuremath{#1~\mathrm{GeV}/c^{2}}}

\newcommand{\invfb}[1]  {\ensuremath{#1~\mathrm{fb}^{-1}}}

\newcommand{\gevccnoarg}{\ensuremath{\mathrm{GeV}/c^{2}}\xspace}

\newcommand{\measErr}[2]{\ensuremath{#1 \pm #2}\xspace}

\newcommand{\measStatSyst}[3]{\ensuremath{#1 \pm #2~(\textrm{stat}) \pm #3~(\textrm{syst})}\xspace}

\begin{document}

\title{Top quark mass measurement using the template method at CDF}

\affiliation{Institute of Physics, Academia Sinica, Taipei, Taiwan 11529, Republic of China} 
\affiliation{Argonne National Laboratory, Argonne, Illinois 60439, USA} 
\affiliation{University of Athens, 157 71 Athens, Greece} 
\affiliation{Institut de Fisica d'Altes Energies, ICREA, Universitat Autonoma de Barcelona, E-08193, Bellaterra (Barcelona), Spain} 
\affiliation{Baylor University, Waco, Texas 76798, USA} 
\affiliation{Istituto Nazionale di Fisica Nucleare Bologna, $^z$University of Bologna, I-40127 Bologna, Italy} 
\affiliation{University of California, Davis, Davis, California 95616, USA} 
\affiliation{University of California, Los Angeles, Los Angeles, California 90024, USA} 
\affiliation{Instituto de Fisica de Cantabria, CSIC-University of Cantabria, 39005 Santander, Spain} 
\affiliation{Carnegie Mellon University, Pittsburgh, Pennsylvania 15213, USA} 
\affiliation{Enrico Fermi Institute, University of Chicago, Chicago, Illinois 60637, USA}
\affiliation{Comenius University, 842 48 Bratislava, Slovakia; Institute of Experimental Physics, 040 01 Kosice, Slovakia} 
\affiliation{Joint Institute for Nuclear Research, RU-141980 Dubna, Russia} 
\affiliation{Duke University, Durham, North Carolina 27708, USA} 
\affiliation{Fermi National Accelerator Laboratory, Batavia, Illinois 60510, USA} 
\affiliation{University of Florida, Gainesville, Florida 32611, USA} 
\affiliation{Laboratori Nazionali di Frascati, Istituto Nazionale di Fisica Nucleare, I-00044 Frascati, Italy} 
\affiliation{University of Geneva, CH-1211 Geneva 4, Switzerland} 
\affiliation{Glasgow University, Glasgow G12 8QQ, United Kingdom} 
\affiliation{Harvard University, Cambridge, Massachusetts 02138, USA} 
\affiliation{Division of High Energy Physics, Department of Physics, University of Helsinki and Helsinki Institute of Physics, FIN-00014, Helsinki, Finland} 
\affiliation{University of Illinois, Urbana, Illinois 61801, USA} 
\affiliation{The Johns Hopkins University, Baltimore, Maryland 21218, USA} 
\affiliation{Institut f\"{u}r Experimentelle Kernphysik, Karlsruhe Institute of Technology, D-76131 Karlsruhe, Germany} 
\affiliation{Center for High Energy Physics: Kyungpook National University, Daegu 702-701, Korea; Seoul National University, Seoul 151-742, Korea; Sungkyunkwan University, Suwon 440-746, Korea; Korea Institute of Science and Technology Information, Daejeon 305-806, Korea; Chonnam National University, Gwangju 500-757, Korea; Chonbuk National University, Jeonju 561-756, Korea} 
\affiliation{Ernest Orlando Lawrence Berkeley National Laboratory, Berkeley, California 94720, USA} 
\affiliation{University of Liverpool, Liverpool L69 7ZE, United Kingdom} 
\affiliation{University College London, London WC1E 6BT, United Kingdom} 
\affiliation{Centro de Investigaciones Energeticas Medioambientales y Tecnologicas, E-28040 Madrid, Spain} 
\affiliation{Massachusetts Institute of Technology, Cambridge, Massachusetts 02139, USA} 
\affiliation{Institute of Particle Physics: McGill University, Montr\'{e}al, Qu\'{e}bec, Canada H3A~2T8; Simon Fraser University, Burnaby, British Columbia, Canada V5A~1S6; University of Toronto, Toronto, Ontario, Canada M5S~1A7; and TRIUMF, Vancouver, British Columbia, Canada V6T~2A3} 
\affiliation{University of Michigan, Ann Arbor, Michigan 48109, USA} 
\affiliation{Michigan State University, East Lansing, Michigan 48824, USA}
\affiliation{Institution for Theoretical and Experimental Physics, ITEP, Moscow 117259, Russia}
\affiliation{University of New Mexico, Albuquerque, New Mexico 87131, USA} 
\affiliation{Northwestern University, Evanston, Illinois 60208, USA} 
\affiliation{The Ohio State University, Columbus, Ohio 43210, USA} 
\affiliation{Okayama University, Okayama 700-8530, Japan} 
\affiliation{Osaka City University, Osaka 588, Japan} 
\affiliation{University of Oxford, Oxford OX1 3RH, United Kingdom} 
\affiliation{Istituto Nazionale di Fisica Nucleare, Sezione di Padova-Trento, $^{aa}$University of Padova, I-35131 Padova, Italy} 
\affiliation{LPNHE, Universite Pierre et Marie Curie/IN2P3-CNRS, UMR7585, Paris, F-75252 France} 
\affiliation{University of Pennsylvania, Philadelphia, Pennsylvania 19104, USA}
\affiliation{Istituto Nazionale di Fisica Nucleare Pisa, $^{bb}$University of Pisa, $^{cc}$University of Siena and $^{dd}$Scuola Normale Superiore, I-56127 Pisa, Italy} 
\affiliation{University of Pittsburgh, Pittsburgh, Pennsylvania 15260, USA} 
\affiliation{Purdue University, West Lafayette, Indiana 47907, USA} 
\affiliation{University of Rochester, Rochester, New York 14627, USA} 
\affiliation{The Rockefeller University, New York, New York 10065, USA} 
\affiliation{Istituto Nazionale di Fisica Nucleare, Sezione di Roma 1, $^{ee}$Sapienza Universit\`{a} di Roma, I-00185 Roma, Italy} 

\affiliation{Rutgers University, Piscataway, New Jersey 08855, USA} 
\affiliation{Texas A\&M University, College Station, Texas 77843, USA} 
\affiliation{Istituto Nazionale di Fisica Nucleare Trieste/Udine, I-34100 Trieste, $^{ff}$University of Trieste/Udine, I-33100 Udine, Italy} 
\affiliation{University of Tsukuba, Tsukuba, Ibaraki 305, Japan} 
\affiliation{Tufts University, Medford, Massachusetts 02155, USA} 
\affiliation{University of Virginia, Charlottesville, VA  22906, USA}
\affiliation{Waseda University, Tokyo 169, Japan} 
\affiliation{Wayne State University, Detroit, Michigan 48201, USA} 
\affiliation{University of Wisconsin, Madison, Wisconsin 53706, USA} 
\affiliation{Yale University, New Haven, Connecticut 06520, USA} 
\author{T.~Aaltonen}
\affiliation{Division of High Energy Physics, Department of Physics, University of Helsinki and Helsinki Institute of Physics, FIN-00014, Helsinki, Finland}
\author{B.~\'{A}lvarez~Gonz\'{a}lez$^v$}
\affiliation{Instituto de Fisica de Cantabria, CSIC-University of Cantabria, 39005 Santander, Spain}
\author{S.~Amerio}
\affiliation{Istituto Nazionale di Fisica Nucleare, Sezione di Padova-Trento, $^{aa}$University of Padova, I-35131 Padova, Italy} 

\author{D.~Amidei}
\affiliation{University of Michigan, Ann Arbor, Michigan 48109, USA}
\author{A.~Anastassov}
\affiliation{Northwestern University, Evanston, Illinois 60208, USA}
\author{A.~Annovi}
\affiliation{Laboratori Nazionali di Frascati, Istituto Nazionale di Fisica Nucleare, I-00044 Frascati, Italy}
\author{J.~Antos}
\affiliation{Comenius University, 842 48 Bratislava, Slovakia; Institute of Experimental Physics, 040 01 Kosice, Slovakia}
\author{G.~Apollinari}
\affiliation{Fermi National Accelerator Laboratory, Batavia, Illinois 60510, USA}
\author{J.A.~Appel}
\affiliation{Fermi National Accelerator Laboratory, Batavia, Illinois 60510, USA}
\author{A.~Apresyan}
\affiliation{Purdue University, West Lafayette, Indiana 47907, USA}
% ...
\collaboration{CDF Collaboration\footnote{With visitors from $^a$University of Massachusetts Amherst, Amherst, Massachusetts 01003,
$^b$Istituto Nazionale di Fisica Nucleare, Sezione di Cagliari, 09042 Monserrato (Cagliari), Italy,
$^c$University of California Irvine, Irvine, CA  92697, 
$^d$University of California Santa Barbara, Santa Barbara, CA 93106
$^e$University of California Santa Cruz, Santa Cruz, CA  95064,
$^f$CERN,CH-1211 Geneva, Switzerland,
$^g$Cornell University, Ithaca, NY  14853, 
$^h$University of Cyprus, Nicosia CY-1678, Cyprus, 
$^i$University College Dublin, Dublin 4, Ireland,
$^j$University of Fukui, Fukui City, Fukui Prefecture, Japan 910-0017,
$^k$Universidad Iberoamericana, Mexico D.F., Mexico,
$^l$Iowa State University, Ames, IA  50011,
$^m$University of Iowa, Iowa City, IA  52242,
$^n$Kinki University, Higashi-Osaka City, Japan 577-8502,
$^o$Kansas State University, Manhattan, KS 66506,
$^p$University of Manchester, Manchester M13 9PL, England,
$^q$Queen Mary, University of London, London, E1 4NS, England,
$^r$Muons, Inc., Batavia, IL 60510,
$^s$Nagasaki Institute of Applied Science, Nagasaki, Japan, 
$^t$National Research Nuclear University, Moscow, Russia,
$^u$University of Notre Dame, Notre Dame, IN 46556,
$^v$Universidad de Oviedo, E-33007 Oviedo, Spain, 
$^w$Texas Tech University, Lubbock, TX  79609, 
$^x$Universidad Tecnica Federico Santa Maria, 110v Valparaiso, Chile,
$^y$Yarmouk University, Irbid 211-63, Jordan,
$^{gg}$On leave from J.~Stefan Institute, Ljubljana, Slovenia, 
}}
\noaffiliation

\date{\today}
\begin{abstract}
We present a measurement of the top quark mass in the lepton+jets and dilepton channels of \ttbar decays using the template method. The data sample corresponds to an integrated luminosity of \invfb{5.6} of \ppbar collisions at Tevatron with $\sqrt{s}=1.96$~TeV, collected with the CDF II detector. The measurement is performed by constructing templates of three kinematic variables in the lepton+jets and two kinematic variables in the dilepton channel. The variables are two reconstructed top quark masses from different jets-to-quarks combinations and the invariant mass of two jets from the $W$ decay in the lepton+jets channel, and a reconstructed top quark mass and $\mTT$, a variable related to the transverse mass in events 
with two missing particles, in the dilepton channel. The simultaneous fit of the templates from signal and background events in the lepton+jets and dilepton channels to the data yields a measured top quark mass of
$\mtop = \gevcc{\measStatSyst{172.1}{1.1}{0.9}}.$

\end{abstract}

\maketitle

The top quark~($t$) is by far the heaviest known elementary particle, with a mass approximately 40 times larger than the mass of its isospin partner, the bottom quark~($b$)~\cite{pdg}. 
The top quark contributes significantly to electroweak radiative 
corrections relating the top quark mass~($\mtop$) and the $W$ boson mass to the mass of the predicted Higgs boson within either the standard model~(SM) or beyond the SM~\cite{elfit,gfit}. 
Precision measurements of $\mtop$ provide therefore important constraints on the Higgs boson mass in either model. 
Since the discovery of the top quark in 1995~\cite{top_discovery} at the Fermilab Tevatron \ppbar Collider, 
both the CDF and D0 experiments have been improving the precision of the $\mtop$ measurement~\cite{top_average}. 
However it is important to measure $\mtop$ using different techniques and independent data samples in different decay channels.
Significant differences in the measurements of $\mtop$ in different decay channels could indicate 
contributions from new physics beyond the SM~\cite{bSM}.

This letter reports a measurement of the top quark mass using the template method~\cite{tmt1,tmt2,tmt3}. 
We use samples of \ttbar candidates in the lepton+jets and dilepton channels, corresponding
to an integrated luminosity of \invfb{5.6} of proton-antiproton collisions at $\sqrt{s}=1.96$~TeV, 
collected by the CDF II detector~\cite{r_cdfDetector}.  
This is a general-purpose detector designed
to study \ppbar collisions at the Fermilab Tevatron.
A charged-particle tracking system, consisting of a silicon
microstrip tracker and a drift chamber, is immersed
in a 1.4~T magnetic field. Electromagnetic and hadronic
calorimeters surround the tracking system and measure
particle energies. Drift chambers and scintillators, located
outside the calorimeters, detect muon candidates.  

Assuming unitarity of the three-generation CKM matrix, the top quark decays almost exclusively into a $W$ boson and a $b$ quark~\cite{pdg}. 
The case where one $W$ decays leptonically into an electron or a muon plus a neutrino and the other 
hadronically into a pair of jets defines the lepton+jets decay channel. 
The dilepton channel is defined as the case where both $W$'s decay leptonically into an electron or a muon plus a neutrino.  

Lepton+jets events are selected by requiring one isolated~\cite{isolated} electron (muon) with 
$\et > \gev{20}$ ($\pt > \gevc{20}$) and pseudorapidity 
$|\eta|<1.1$~\cite{cdfco}. 
We also require high missing transverse energy~\cite{met}, $\met > \gev{20}$, and at least four jets. 
Jets are reconstructed with a cone algorithm~\cite{jetclu} with radius 
$R = \sqrt{(\Delta \eta)^2+(\Delta \phi)^2} = 0.4$. 
Jets originating from $b$ quarks are identified~(tagged) using a secondary vertex tagging algorithm~\cite{secvtx}. 
We request at least one jet to be tagged as a $b$ jet. 
We divide the sample of candidate lepton+jets events into sub-samples of 
one $b$-tagged jet~(1-tag) and two or more $b$-tagged jets~(2-tag). In events with more than two $b$-tagged jets, we consider the two highest $E_{T}$ jets as $b$ quark candidates and treat the other $b$-tagged jets as non $b$-tagged jets. 
In the 1-tag sample, we require exactly four jets 
with transverse energy $\et > \gev{20}$ and $|\eta|<2.0$. In the 2-tag sample,
three jets are required to have $\et > \gev{20}$ and $|\eta| <2.0$, and at least one more jet is required to have $\et > \gev{12}$ and $|\eta|<2.4$.
We apply an additional cut on the scalar sum of transverse energies 
in the event, $\Ht = \etl + \met + \sum_{jets}\etj$, requiring $\Ht > \gev{250}$ for all events to further reject backgrounds. $\etm = \ptm$ is assumed in the \Ht calculation.

The primary sources of background in the lepton+jets channel are $W$+jets and QCD multijet production. 
We also consider small contributions from $Z$+jets, diboson, and single-top production. 
To estimate the contribution of each process, we use a combination of data and Monte Carlo~(MC) based 
techniques described in Ref.~\cite{xsection}. For the $Z$+jets, diboson, single top, and $\ttbar$ events
we normalize MC simulation events using their respective theoretical cross sections~\cite{smcross1,smcross2,thcross}. QCD multijet background is estimated using 
the data referring to techniques described in Ref.~\cite{qcdback}. The shape of $W$+jets background is obtained from MC while the number of $W$+jets 
events is determined from the data by subtracting all the other backgrounds and $\ttbar$. 

Three observables are used from each lepton+jets event: two reconstructed top quark masses~(\mtr and \mttr) and the invariant mass of 
the two jets from the hadronically decaying $W$ boson~($\mjj$).  We have complete reconstruction of the \ttbar kinematics in the lepton+jets channel~\cite{tmt1,tmt2} with constraints from the precise $W$ boson mass and requiring the $t$ and $\bar{t}$ masses to be the same. 
With the assumption that the leading four jets in the 
detector come from the \ttbar decay products, there are six and two possible assignments 
of jets to quarks for 1-tag and 2-tag respectively.
A minimization is performed for each assignment using a $\chi^2$ comparison to the \ttbar hypothesis with \mtr taken from the assignment that yields 
the lowest $\chi^2$. To increase the statistical power of the measurement, we employ an additional
observable \mttr from the assignment that yields the 2$^{nd}$ lowest $\chi^{2}$. 
Events with the lowest $\chi^2 > 9.0$ are removed from the sample to reject poorly reconstructed events. 
The dijet mass \mjj is calculated as the invariant mass of two non $b$-tagged jets which provides 
the closest value to the world average $W$ boson mass of \gevcc{80.40}~\cite{pdg}. 
We apply boundary cuts on \mtr and \mttr~($\gevcc{100} < \mtr, \mttr < \gevcc{350}$) and 
$\mjj$~($\gevcc{50} < \mjj < \gevcc{120}$ for 1-tag events and 
$\gevcc{50} < \mjj < \gevcc{125}$ for 2-tag events), and normalize the probability 
density function in the signal region. 
The estimated number of background events and the observed number of events 
after event selection, $\chi^2$ cut, and boundary cuts are 
listed in Table~\ref{table_background_lj} for the lepton+jets decay channel.

\begin{table}
\begin{center}
\caption[Expected number of background and signal events in the lepton+jets channel]{
Expected and observed numbers of signal and background events assuming \ttbar production 
cross section $\sigma_{\ttbar}$ = \pb{7.4} and \mtop = \gevcc{172.5} 
in the lepton+jets channel. 
}
\begin{ruledtabular}
\label{table_background_lj}
\begin{tabular}{lcc}
             &1-tag       & 2-tag\\
\hline
$W$+jets       &53.4 $\pm$ 17.5 &8.5 $\pm$ 3.0\\
QCD multijet &13.1 $\pm$ 10.6 & 1.8 $\pm$ 1.5\\
$Z$+jets       &4.7 $\pm$ 1.0  & 0.5 $\pm$ 0.1\\
Diboson      &6.3 $\pm$ 0.8  & 0.8 $\pm$ 0.1\\
Single top   &4.9 $\pm$ 0.4  & 2.0 $\pm$ 0.2\\
\hline
Background   &105 $\pm$ 21 & 14.2 $\pm$ 3.3 \\
\ttbar signal&590 $\pm$ 74 & 293 $\pm$ 45\\
\hline
Expected& 694 $\pm$ 77  & 307 $\pm$ 45 \\
Observed     &695 & 286 \\
\end{tabular}
\end{ruledtabular}
\end{center}
\end{table}

To select dilepton candidate events, we require two oppositely charged leptons with $\et > \gev{20}$ (for electrons) or
$\pt > \gevc{20}$ (for muons).
One lepton is required to be isolated in the central region~($|\eta| < 1.1$) 
of the detector, but the other can be a non-isolated lepton in the central region or an isolated electron in the forward region~($1.1 < |\eta| < 2.0$).
We also require $\met > \gev{25}$, and at least two jets with $\et > \gev{15}$ and 
$|\eta| < 2.5$. To further reject backgrounds, we require $\Ht > \gev{200}$. 
In measuring the top quark mass, we divide the dilepton sample into events with $b$-tagged jets~(tagged) 
and without $b$-tagged jets~(non-tagged).

Drell-Yan, diboson, and $W$+jets~(fake lepton) events are the primary sources of background in the dilepton channel.
We estimate the rate of the Drell-Yan and diboson events with calculations based on MC simulations. 
For the Drell-Yan $Z$+jets process, we normalize the MC sample by matching the number of $Z$ events 
predicted and observed in the $Z$ mass region between \gevcc{76} and \gevcc{106}. We use data to estimate 
the rate of $W$+jets~(fake lepton) events where an event has one real lepton and one of the jets misidentified 
as the other lepton. The detailed procedure of background estimation in the dilepton channel is described in Ref.~\cite{dilxsection}. 
For each event we calculate a reconstructed top quark mass \mtnwa using the neutrino weighting 
algorithm~\cite{nwa}, and we calculate a quantity \mTT~\cite{mT2}. Here \mTT is a variable related to the transverse mass of the mother particles~(top quark in the \ttbar system) in events 
with two missing particles from pair production of the mother particles.
We firstly use this variable for the top quark mass measurement in the dilepton channel~\cite{tmt3}. 
We require these observables to be consistent with the top quark signal
by demanding $\gevcc{100} < \mtnwa < \gevcc{350}$ and $\gevcc{30} < \mTT < \gevcc{200}$.
The estimated number of background events and the observed number of events after event selection 
are listed in Table~\ref{table_background_dil} for the dilepton decay channel. 

\begin{table}
\begin{center}
\caption[Expected number of background and signal events in the dilepton channel]{
Expected and observed number of signal and background events 
assuming \ttbar production cross section $\sigma_{\ttbar}$ = \pb{7.4} and \mtop = \gevcc{172.5} 
in the dilepton channel. 
}
\begin{ruledtabular}
\label{table_background_dil}
\begin{tabular}{lcc}
             &non-tagged       &tagged\\
\hline
Diboson       &19.2 $\pm$ 3.3  & 0.7 $\pm$ 0.2\\
Drell-Yan     &31.5 $\pm$ 3.9  & 3.7 $\pm$ 0.2\\
$W$+jets~(fake lepton) &30.8 $\pm$ 9.4  & 4.6 $\pm$ 1.3\\
\hline
Background   &81.6 $\pm$ 10.4 & 8.9 $\pm$ 1.4 \\
\ttbar signal&124 $\pm$ 16 & 151 $\pm$ 19\\
\hline
Expected& 205 $\pm$ 19  & 160 $\pm$ 19 \\
Observed     &237 & 155 \\
\end{tabular}
\end{ruledtabular}
\end{center}
\end{table}

We estimate the probability density functions~(p.d.f.'s) of signal and
background using kernel density
estimation~(KDE)~\cite{KDEHEP} that constructs the p.d.f. without 
any assumption of a functional form. 
In the lepton+jets channel, we use the three dimensional KDE that accounts for the correlation between the three observables. In the dilepton channel, instead, we use the two dimensional KDE. 
The dijet mass \mjj of the two jets assigned to the $W$ in the lepton+jets channel is used for {\it in situ} calibration of jet energy scale~(JES)~\cite{tmt1,tmt2}. 
The p.d.f.'s for the observables are estimated at discrete values of \mtop from \gevcc{130} to \gevcc{220}, with increments from \gevcc{0.5} in the region immediately above and below \gevcc{172.5} to \gevcc{5} near the extreme mass values, and at discrete values of \djes from $-$3.0~$\sigma_{c}$ to 3.0~$\sigma_c$ with increments of 0.2~$\sigma_c$, where $\sigma_c$ is the CDF JES fractional uncertainty~\cite{jes} and \djes corresponds to the difference between the energy scale in MC simulation and data. 
We interpolate the MC distributions 
to find p.d.f.'s for arbitrary values of \mtop and \djes using the local polynomial smoothing
method~\cite{lps}.  
We fit the signal and background p.d.f.'s to the distributions of the observables in the data using
an unbinned maximum likelihood fit~\cite{eml} where
we minimize the negative logarithm of the likelihood using {\sc minuit}~\cite{minuit}. 
The likelihood is built for each sub-sample separately, 1-tag and 2-tag for lepton+jets events, non-tagged and tagged for dilepton events, and an overall likelihood is then obtained by multiplying them together.
We independently obtain the results from the lepton+jets channel, the dilepton channel, and the two channels combined. 
In the combined fit, the dilepton channel uses the JES calibration found in the lepton+jets channel.
We evaluate the statistical uncertainty on \mtop by searching for the points where the
negative logarithm of the likelihood exeeds the minimum by 0.5.
Ref.~\cite{tmt2,tmt3} provides detailed information about this technique.

We test the mass fit procedures using 3000 pseudoexperiments 
for a set of 14 different \mtop values ranging from \gevcc{159} to \gevcc{185}. 
In each experiment, we select the number of background events from a
Poisson distribution with a mean equal to the expected total number of background events 
in the sample and the number of signal events from a Poisson distribution
with a mean equal to the expected number of signal events normalized to a $\ttbar$ pair 
production cross section of 7.4~pb at \mtop = \gevcc{172.5}~\cite{thcross}. 
The distributions of the average mass residual~(deviation from the
input top mass) and the width of the pull~(the ratio of the residual to the
uncertainty reported by {\sc minuit}) for simulated
experiments are corrected to be unity and zero respectively. The corrections are $m_{corr} = 1.04\times m_{meas}-\gevcc{6.8}$, $m_{corr} = 1.03\times m_{meas}-\gevcc{5.5}$, and $m_{corr} = 1.03\times m_{meas}-\gevcc{5.9}$ for combined fit, lepton+jets, and dilepton channel respectively, where $m_{meas}$ is the raw value from likelihood fit and $m_{corr}$ is the corrected value of the measurement. We increase the measured uncertainty by 4\% for combined fit and lepton+jets channel and 3\% for dilepton channel to correct the width of the pull.

\begin{table}
\begin{center}
\caption[Summary of systematics]{Estimated
systematic uncertainties in the combined fit~(Comb), lepton+jets~(LJ), and dilepton~(DIL) (unit in \gevccnoarg).}
\begin{ruledtabular}
\label{systtablesummary}
\begin{tabular}{lccc}
Source              & Comb & LJ & DIL \\
\hline
(Residual) Jet Energy Scale        & 0.5& 0.5& 3.0 \\
Signal modeling                    & 0.7& 0.7& 0.3 \\
$b$ Jet energy scale               & 0.3& 0.3& 0.4 \\
Initial and final state radiation  & 0.1& 0.1& 0.2 \\
Parton distribution functions      & 0.1& 0.1& 0.3 \\
Gluon fusion fraction              & $<$0.1& $<$0.1& 0.1 \\
Lepton energy                      & $<$0.1& 0.1& 0.3 \\
Background shape                   & 0.1& 0.1& 0.3 \\
Multiple hadron interaction        & 0.2& 0.1& 0.2 \\
Color reconnection                 & 0.2& 0.2& 0.5 \\
MC statistics                      & 0.1& 0.1& 0.1 \\
\hline
Total systematic uncertainty       & 0.9& 0.9& 3.1 \\
\end{tabular}
\end{ruledtabular}
\end{center}
\end{table}

\begin{cfigure1c}
\begin{tabular}{cc}
\includegraphics[width=0.45\textwidth]{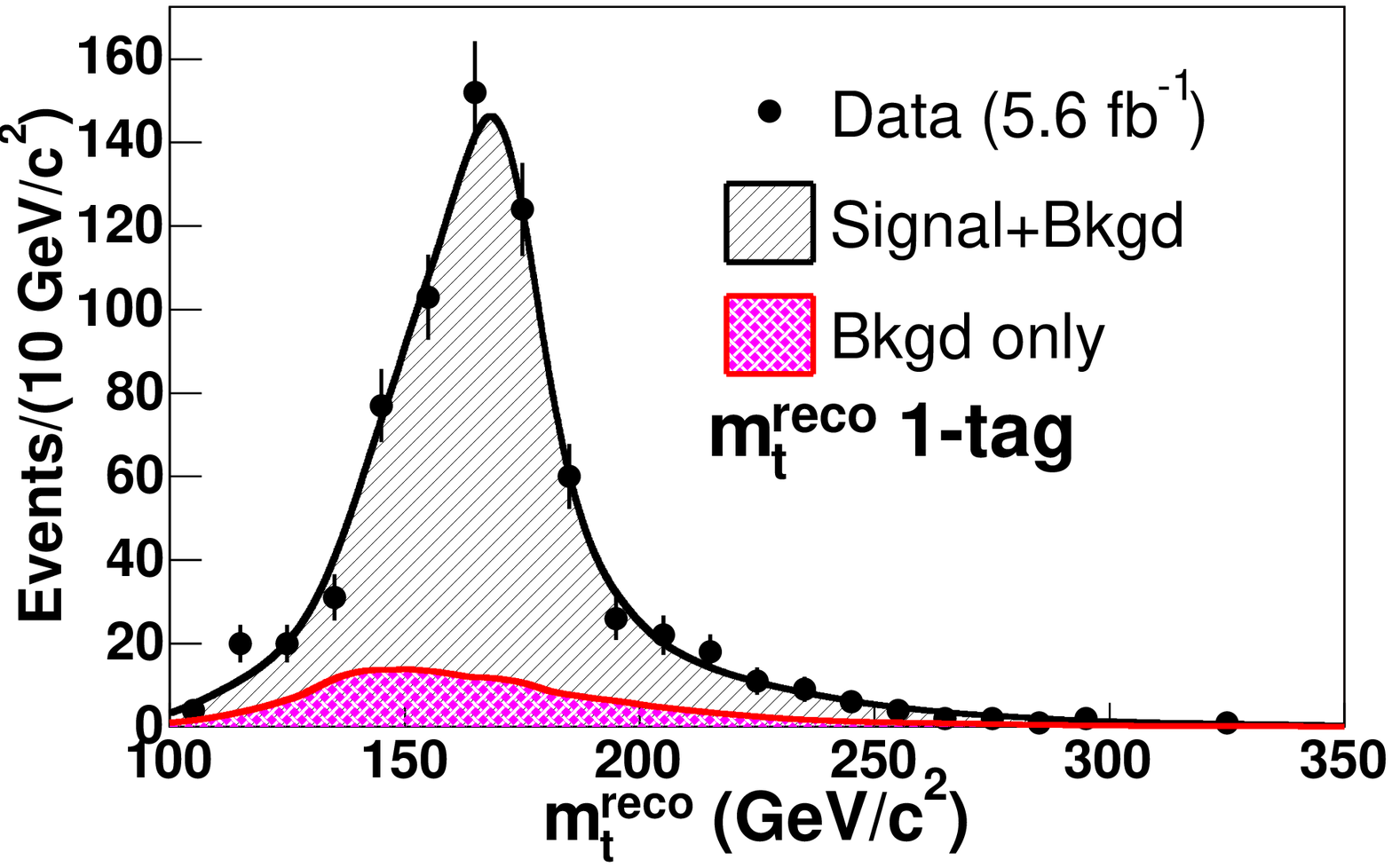}&
\includegraphics[width=0.45\textwidth]{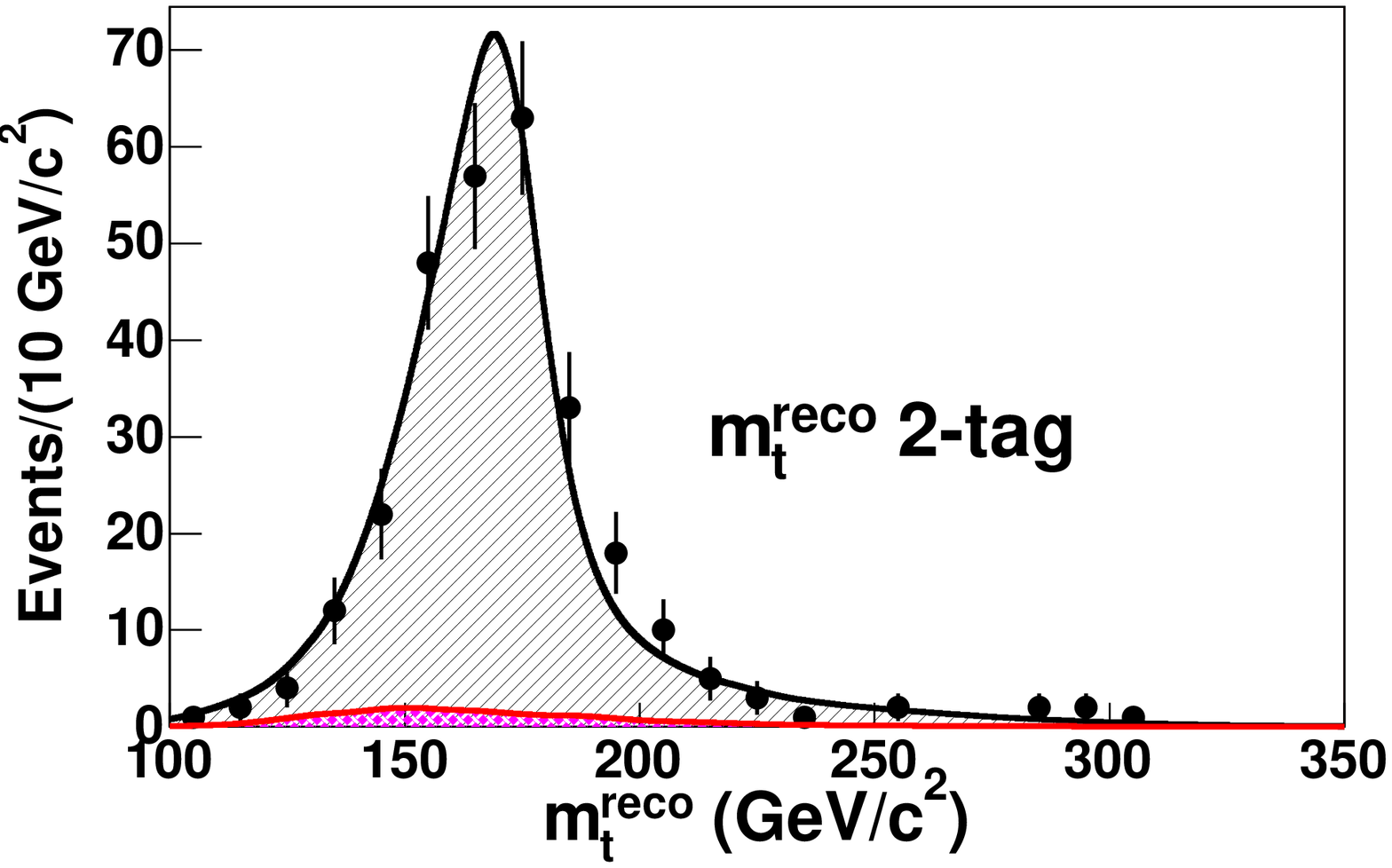}\\
\includegraphics[width=0.45\textwidth]{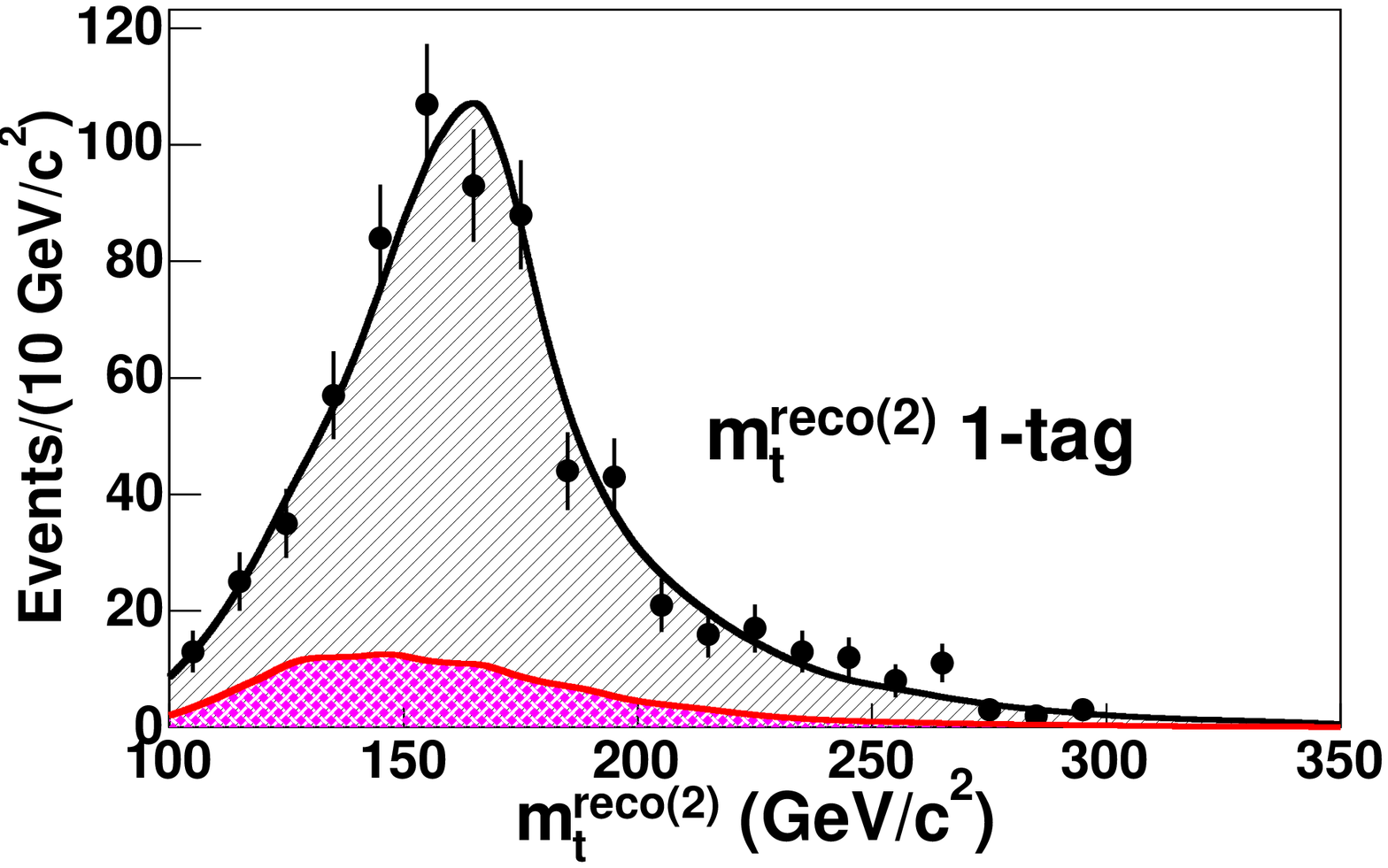}&
\includegraphics[width=0.45\textwidth]{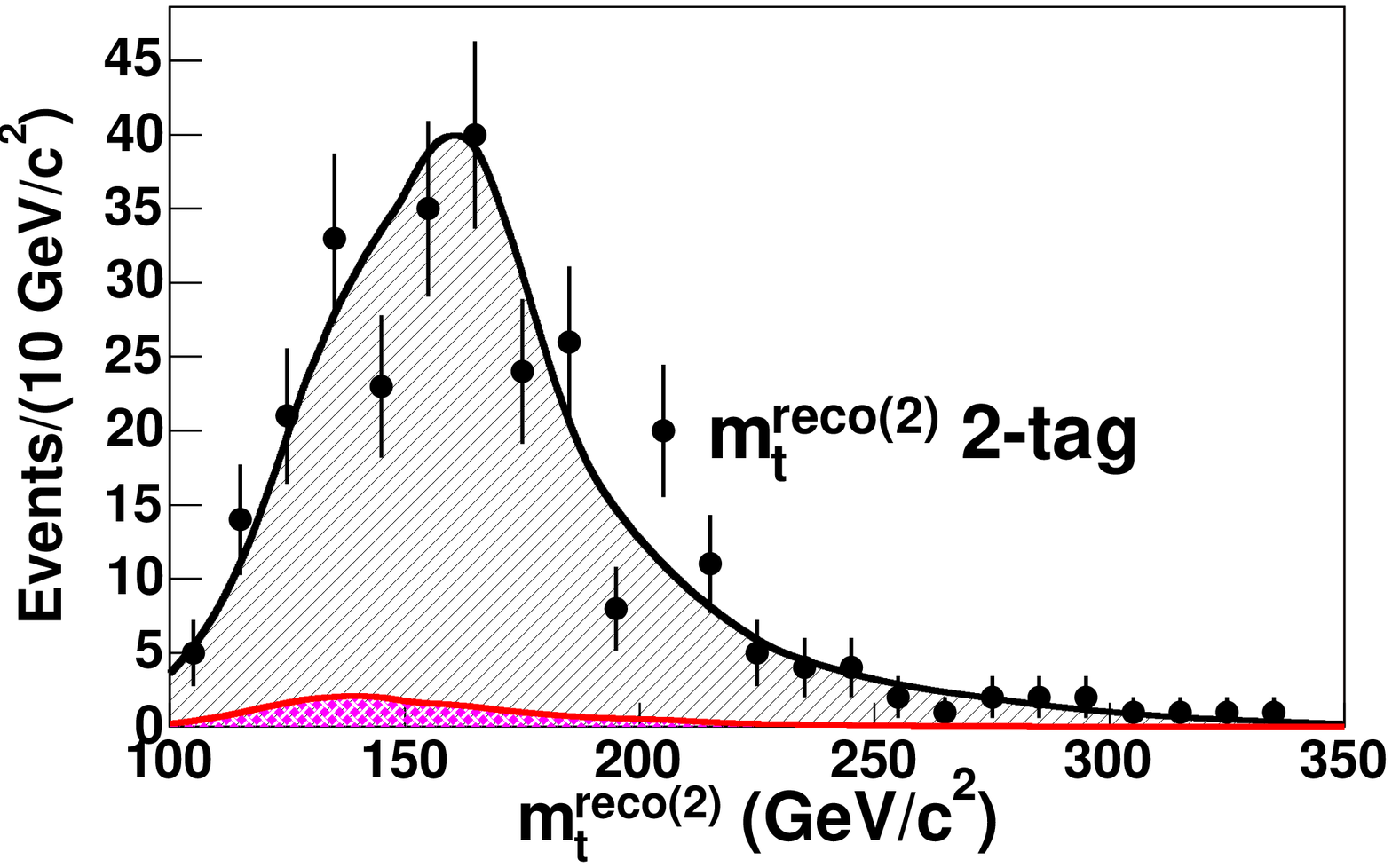}\\
\includegraphics[width=0.45\textwidth]{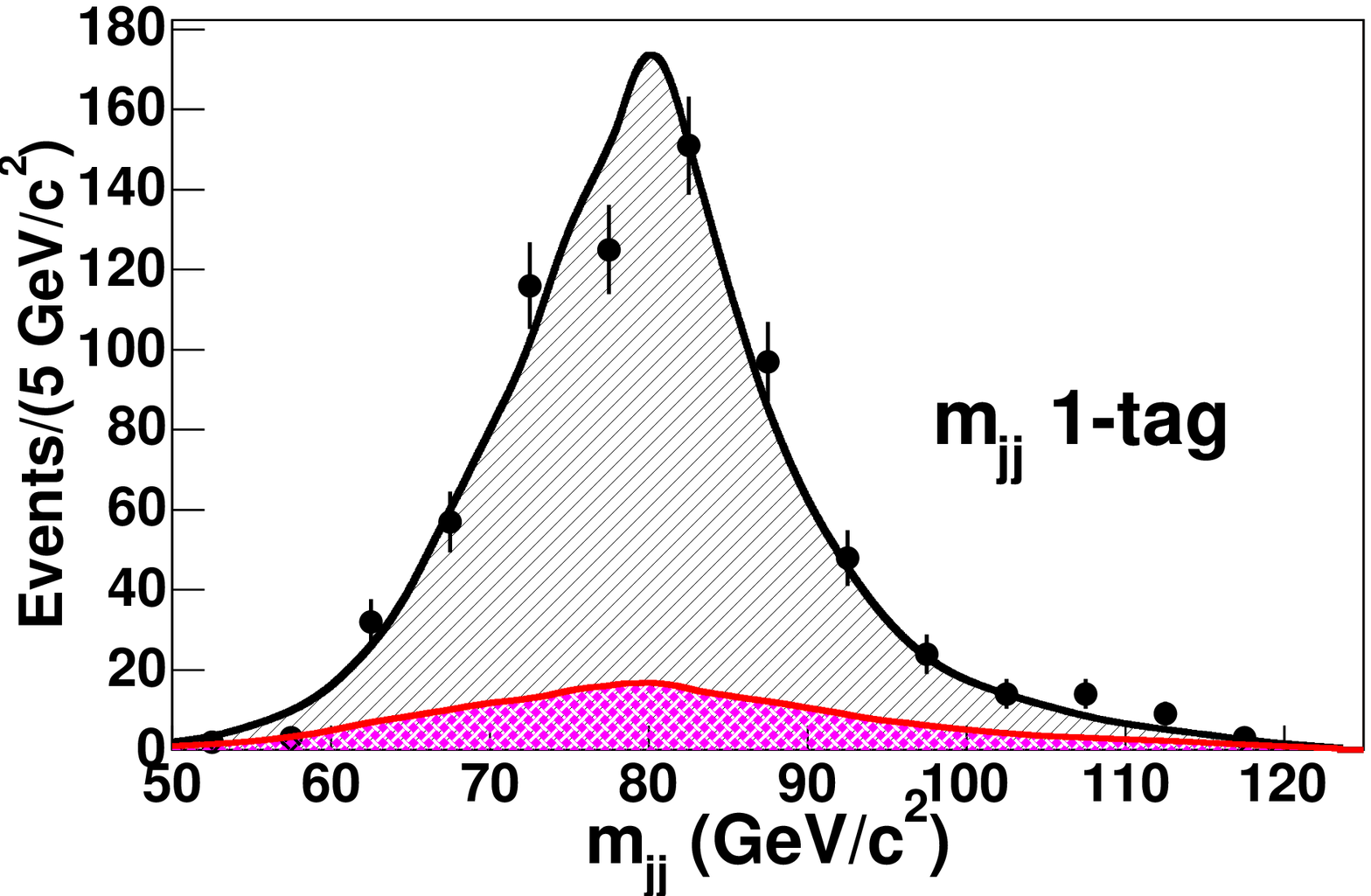}&
\includegraphics[width=0.45\textwidth]{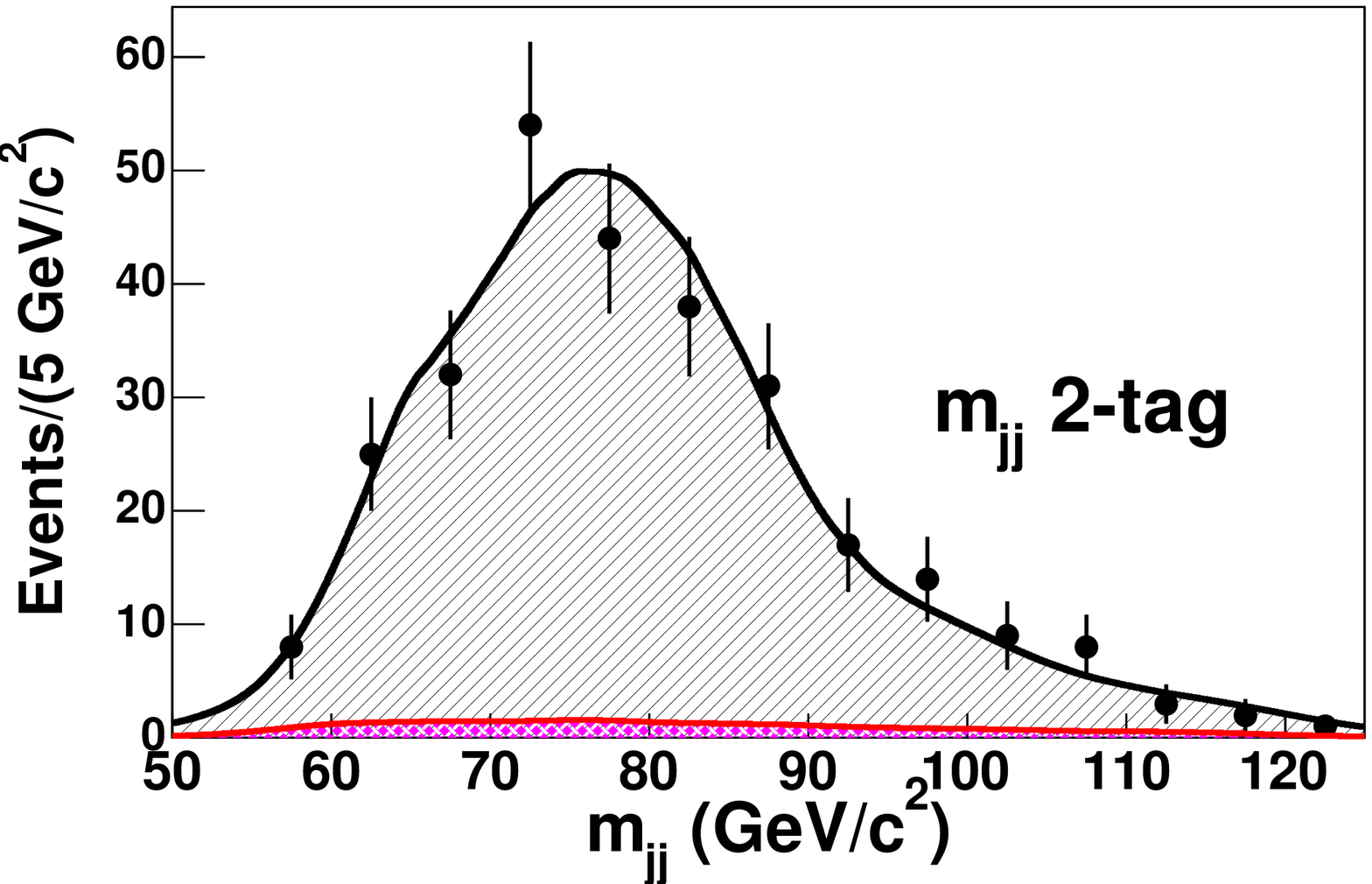}\\
\end{tabular}
\caption[Data fit]{Distributions of the three variables used to measure \mtop in the lepton+jets channel, 
showing 1-tag and 2-tag samples separately. The data are overlaid with the 
predictions from the KDE probability distributions using \mtop = \gevcc{172.0} 
and the full background model. 
}
\label{ref:fit_lj}
\end{cfigure1c}

\begin{cfigure1c}
\begin{tabular}{cc}
\includegraphics[width=0.45\textwidth]{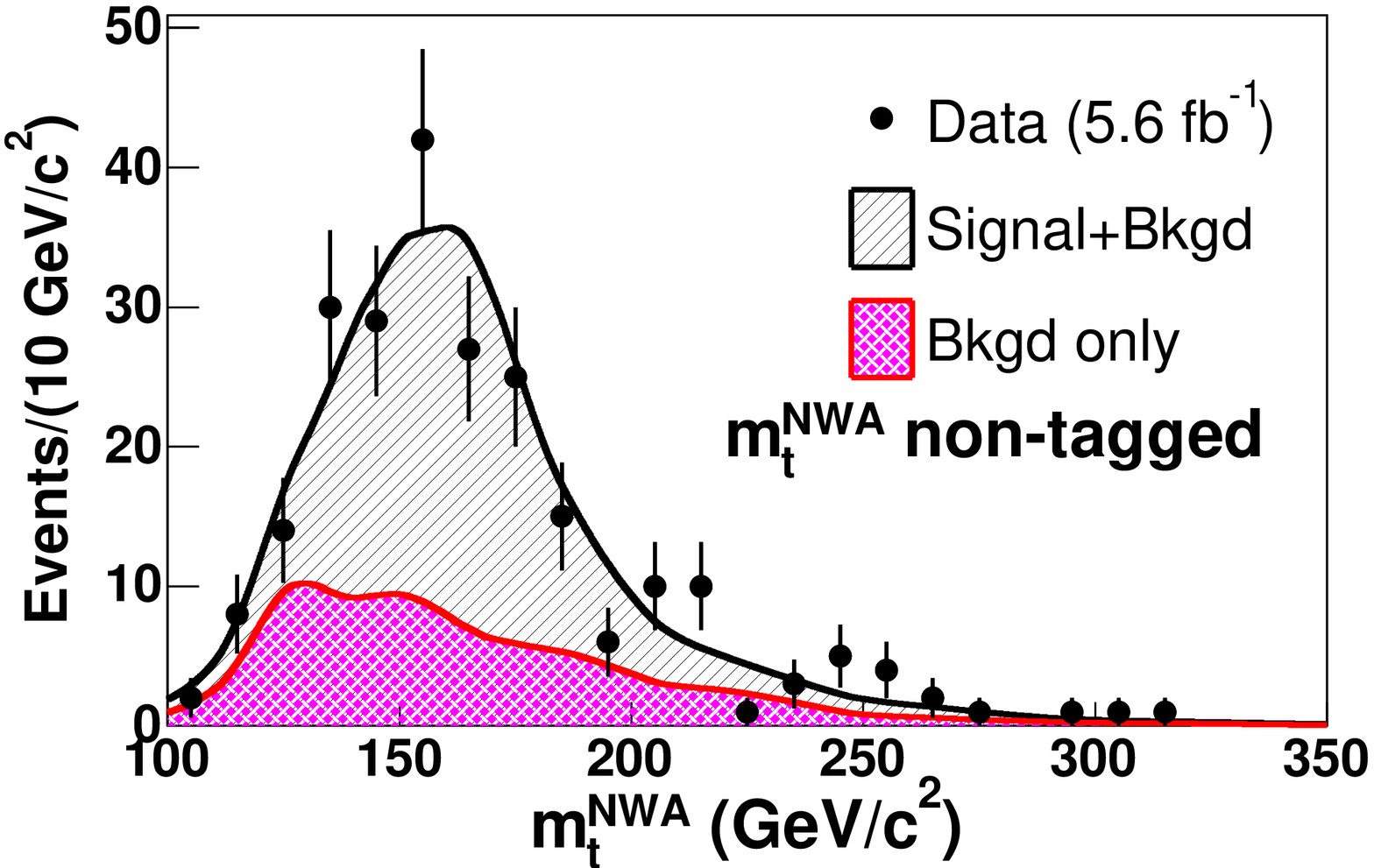}&
\includegraphics[width=0.45\textwidth]{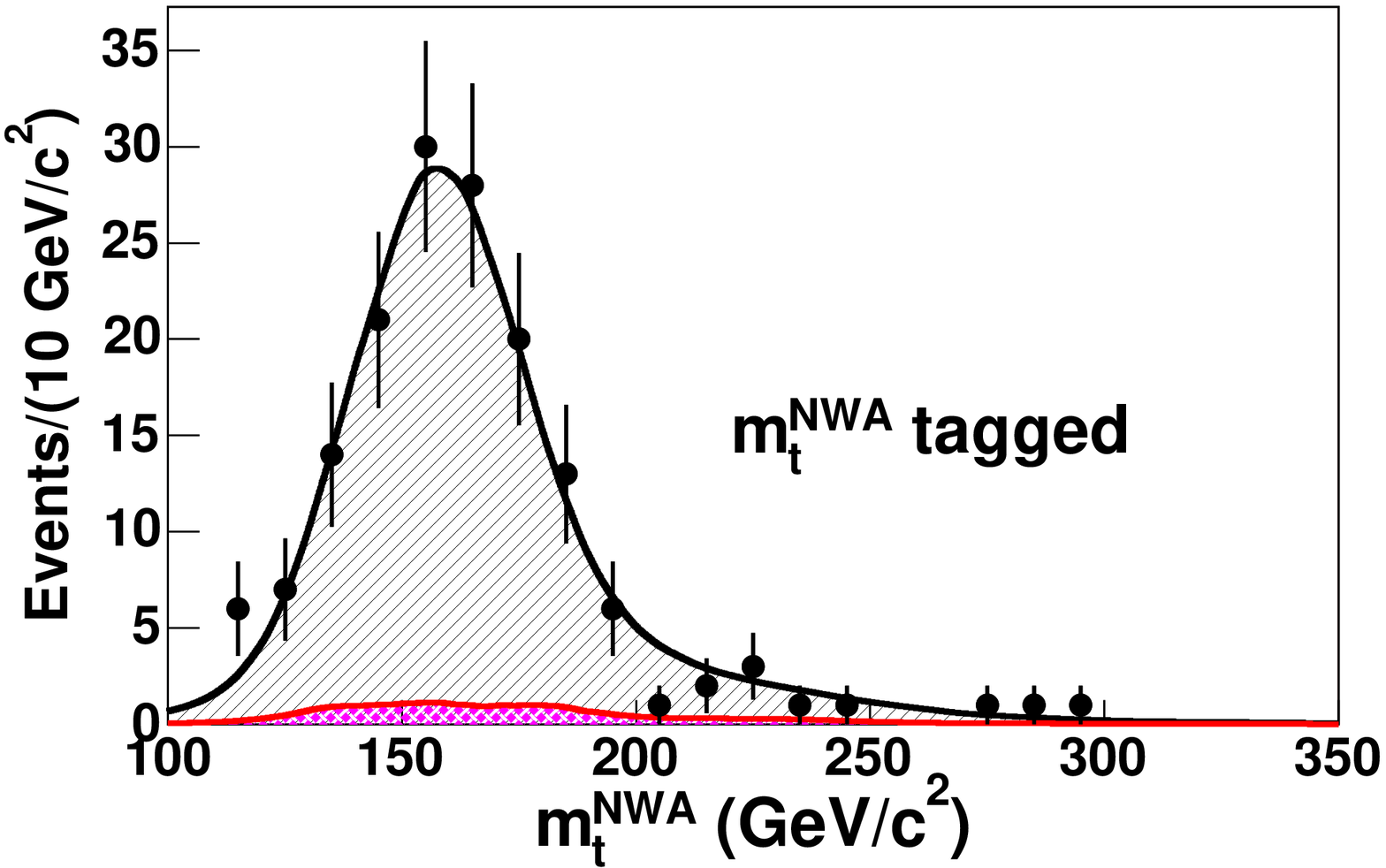}\\
\includegraphics[width=0.45\textwidth]{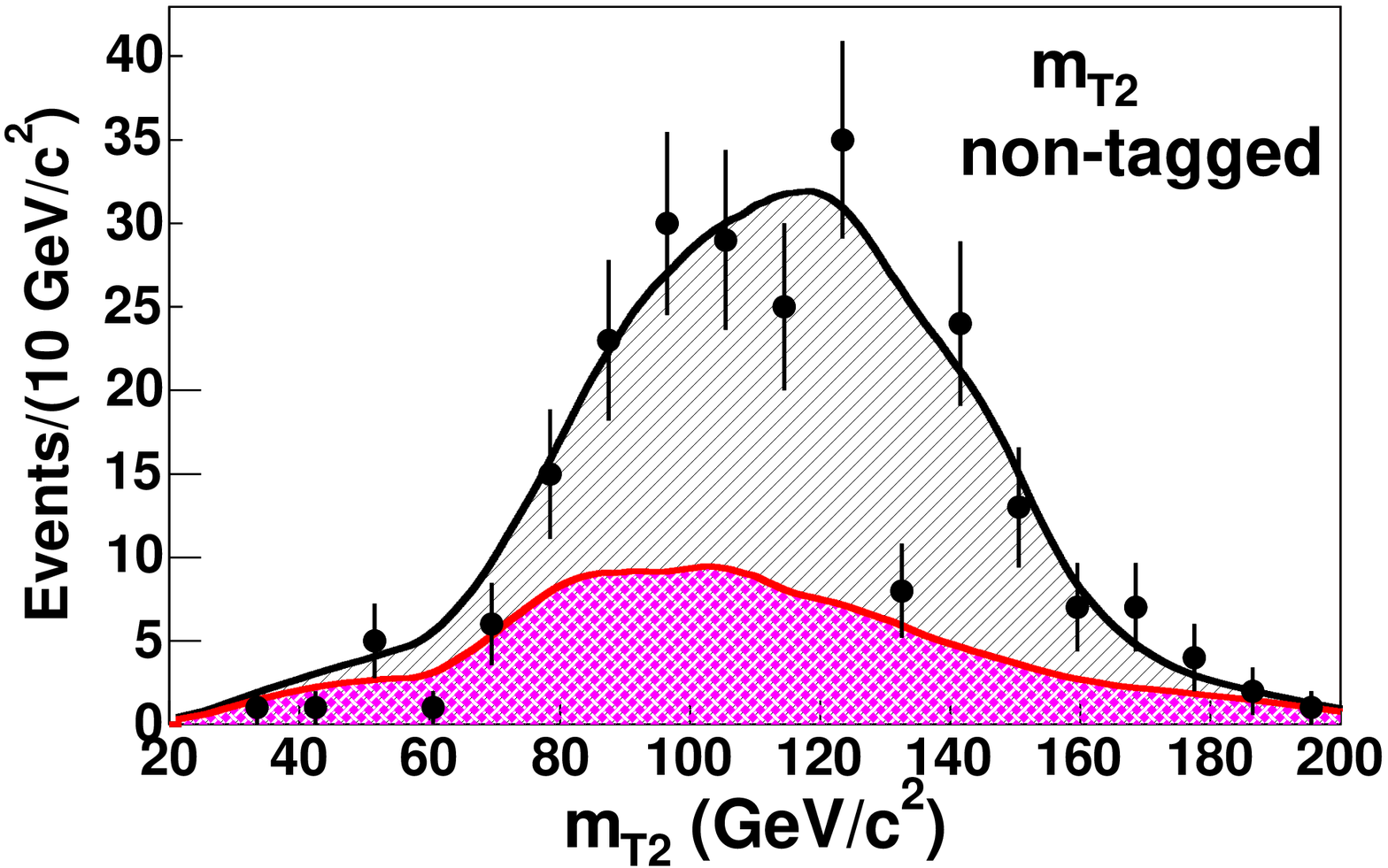}&
\includegraphics[width=0.45\textwidth]{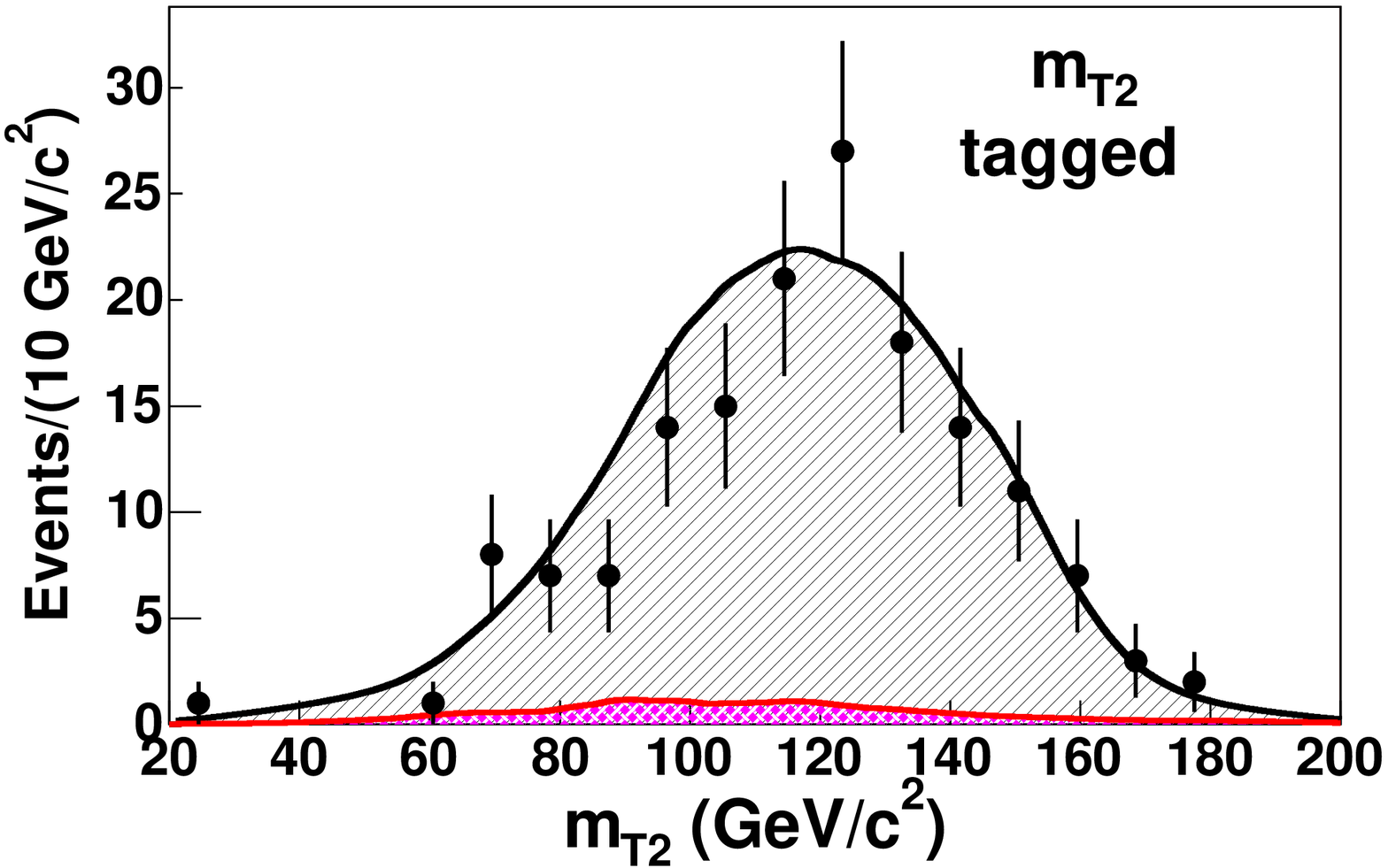}\\
\end{tabular}
\caption[Data fit]{Distributions of the two variables used to measure \mtop 
in the dilepton channel, showing non-tagged and tagged samples separately. The data 
are overlaid with the predictions from the KDE probability distributions using 
\mtop = \gevcc{170.0} and the full background model. 
}
\label{ref:fit_dil}
\end{cfigure1c}

We examine various sources of systematic uncertainties 
that could affect the measurement
by comparing the results of pseudoexperiments in which we vary relevant parameters 
within their systematic uncertainties. The dominant sources of
systematic uncertainty are the residual JES~\cite{tmt2,jes} and signal modeling. 
We vary JES parameters within their uncertainties in both signal and background MC 
generated events and interpret the shifts in the results of the pseudoexperiments as uncertainties. For the dilepton channel, 
which has no {\it in situ} calibration, the JES is the single dominant uncertainty.  
The uncertainty arising from the choice of MC generator~(signal modeling) is estimated
by comparing the results of pseudoexperiments generated with 
{\sc pythia}~\cite{pythia} and {\sc herwig}~\cite{herwig}. 
The $b$-JES systematic uncertainty arising from our modeling of $b$ fragmentation, 
$b$ hadron branching fractions, and calorimeter response captures the additional 
uncertainty not taken into account in the (residual) JES.  We estimate the 
systematic uncertainty due to modeling of initial-state gluon
radiation and final-state gluon radiation by extrapolating
uncertainties in the $p_{T}$ of Drell-Yan events to the \ttbar mass region~\cite{tmt1}.  
We estimate the systematic uncertainty due to parton distribution functions~(PDFs) by varying the independent eigenvectors of the {\sc cteq6m}~\cite{cteq} PDFs, varying $\Lambda_{QCD}$, and comparing the results obtained with {\sc cteq5l}~\cite{cteq5l} and {\sc mrst72}~\cite{mrst} PDFs. 
In estimating the systematic 
uncertainty from the top quark production mechanism, we vary the fraction of top quarks 
produced by gluon-gluon annihilation from 6\% to 20\%, corresponding to the one standard deviation upper bound on the gluon fusion fraction~\cite{gg}. We estimate systematic uncertainties due to 
the lepton energy and momentum scales by propagating shifts in electron energy and muon 
momentum scales within their uncertainties. Background shape systematic uncertainties 
account for the variation of the background composition. We estimate the multiple hadron
interaction systematic uncertainty to account the effect from the difference in 
the average number of interactions between MC samples and the data. 
The color reconnection~(CR) systematic uncertainty~\cite{CR} is evaluated by MC samples generated with and without CR effects using different tunes~\cite{CRtune} of {\sc pythia}. 
The total systematic uncertainties, adding individual components in quadrature, 
are \gevcc{0.9} in the combined fit, \gevcc{0.9} in the lepton+jets channel, 
and \gevcc{3.1} in the dilepton channel. 

We perform the likelihood fits to the data using the observables discussed in this letter and apply the corrections obtained using the simulated experiments. 
We obtain for the lepton+jets channel, a top quark mass
\begin{linenomath*}
\begin{eqnarray*}
\mtop & =& \gevcc{\measStatSyst{172.2}{1.2}{0.9}}\\ 
     & = &\gevcc{\measErr{172.2}{1.5}},
\end{eqnarray*}
\end{linenomath*}
while for the dilepton channel,
\begin{linenomath*}
\begin{eqnarray*}
\mtop & =& \gevcc{\measStatSyst{170.3}{2.0}{3.1}}\\ 
     & = &\gevcc{\measErr{170.3}{3.7}}.
\end{eqnarray*}
\end{linenomath*}
The two channel combined fit yields a top quark mass
\begin{linenomath*}
\begin{eqnarray*}
\mtop & = & \gevcc{\measStatSyst{172.1}{1.1}{0.9}}\\ 
     & = &\gevcc{\measErr{172.1}{1.4}}.
\end{eqnarray*}
\end{linenomath*}
Figure~\ref{ref:fit_lj} shows the measured distributions of the observables 
used for the \mtop measurement in the lepton+jets channel overlaid with density estimates using \ttbar 
signal events with \mtop = \gevcc{172}~(close to the measured \mtop in the lepton+jets channel) and the full background model. 
Figure~\ref{ref:fit_dil} shows the 
corresponding distributions in the dilepton channel using \ttbar signal events with \mtop = \gevcc{170}~(close to the measured \mtop in the dilepton channel). 

In conclusion, we have performed a measurement of the top quark mass using the template method simultaneously in the lepton+jets and dilepton channels. 
The result, $\mtop = \gevcc{\measErr{172.1}{1.4}}$, is consistent with the most recent world average of $\mtop = \gevcc{173.3 \pm 1.1}$~\cite{top_average}. 
In the lepton+jets channel, we use the same data set as the best single \mtop 
measurement~\cite{mtm}, and have a consistent result with slightly larger uncertainty. 
In the dilepton channel, we achieve the single most precise measurement of \mtop in this channel to date
and the result is in good agreement with the measurement in the lepton+jets channel. 

\begin{acknowledgments}
We thank the Fermilab staff and the technical staffs of the participating institutions for their vital contributions. This work was supported by the U.S. Department of Energy and National Science Foundation; the Italian Istituto Nazionale di Fisica Nucleare; the Ministry of Education, Culture, Sports, Science and Technology of Japan; the Natural Sciences and Engineering Research Council of Canada; the National Science Council of the Republic of China; the Swiss National Science Foundation; the A.P. Sloan Foundation; the Bundesministerium f\"ur Bildung und Forschung, Germany; the Korean World Class University Program, the National Research Foundation of Korea; the Science and Technology Facilities Council and the Royal Society, UK; the Institut National de Physique Nucleaire et Physique des Particules/CNRS; the Russian Foundation for Basic Research; the Ministerio de Ciencia e Innovaci\'{o}n, and Programa Consolider-Ingenio 2010, Spain; the Slovak R\&D Agency; the Academy of Finland; and the Australian Research Council (ARC).
\end{acknowledgments}


\begin{thebibliography}{99}
\bibitem{isolated} A lepton is isolated when $\left( \ptt - \ptl \right)/\ptl < 0.1$, where $\ptt$ is the total transverse momentum (energy) and $\ptl$ is the lepton transverse momentum (energy) for muon (electron) in a
  cone of radius $\Delta R \equiv \sqrt{(\Delta\eta)^{2}+(\Delta\phi)^{2}} = 0.4$~\cite{cdfco} with axis along the direction of the lepton.
\bibitem{cdfco} We use a right-handed spherical coordinate system with the origin at the center of the detector with the z-axis along the proton beam and the y-axis poiting up. $\theta$ and $\phi$ are the polar and azimuthal angles, respectively.  The pseudorapidity is defined by $\eta = -\ln \tan(\theta/2)$.  The transverse momentum and energy of a detected particle or jet are defined by 
$\pt = p \sin(\theta)$ and $\et = E \sin(\theta)$, respectively, where $p$ and $E$ are the momentum and energy of the particle. For the reconstructed top quark decay products used in the \mTT calculation, the transverse energy is defined by $\et = \sqrt{m^{2}+\pt^{2}}$, where $m$ is the mass of the product.
\bibitem{met} The missing transverse energy, an imbalance of energy in the transverse plane of the detector, is defined by 
$\met = \vert \sum_{towers}\et{\hat n_{T}} \vert$, where ${\hat n_{T}}$ is the unit vector normal to the beam and pointing to a given calorimeter tower and $\et$ is the transverse energy measured in that tower.
\end{thebibliography}
\end{document}